\documentstyle[epsfig,12pt,a4p]{article}

\newcommand{\etapipi}{\mbox{$\eta \pi^+\pi^-$ }}
\newcommand{\etaeta}{\mbox{$\eta \eta$ }}
\newcommand{\etaetap}{\mbox{$\eta \eta^\prime$ }}
\newcommand{\etapetap}{\mbox{$\eta^\prime \eta^\prime$ }}

\begin{document}
\begin{titlepage}
\def\footnoterule{\hrule width 1.0\columnwidth}
%\hfill  \hfill
%6\thinspace February\thinspace 1990
% \begin{center} {\large EUROPEAN ORGANIZATION FOR NUCLEAR RESEARCH}
%  \end{center}
\begin{tabbing}
put this on the right hand corner using tabbing so it looks
 and neat and in \= \kill
\> {25 November 1999}
\end{tabbing}
\bigskip
\bigskip
\begin{center}{\Large  {\bf A study of the
\etaetap and \etapetap channels produced in central
pp interactions at 450 GeV/c}
}\end{center}

\bigskip
\bigskip
\begin{center}{        The WA102 Collaboration
}\end{center}\bigskip
\begin{center}{
D.\thinspace Barberis$^{  4}$,
F.G.\thinspace Binon$^{   6}$,
F.E.\thinspace Close$^{  3,4}$,
K.M.\thinspace Danielsen$^{ 11}$,
S.V.\thinspace Donskov$^{  5}$,
B.C.\thinspace Earl$^{  3}$,
D.\thinspace Evans$^{  3}$,
B.R.\thinspace French$^{  4}$,
T.\thinspace Hino$^{ 12}$,
S.\thinspace Inaba$^{   8}$,
A.\thinspace Jacholkowski$^{   4}$,
T.\thinspace Jacobsen$^{  11}$,
G.V.\thinspace Khaustov$^{  5}$,
J.B.\thinspace Kinson$^{   3}$,
A.\thinspace Kirk$^{   3}$,
A.A.\thinspace Kondashov$^{  5}$,
A.A.\thinspace Lednev$^{  5}$,
V.\thinspace Lenti$^{  4}$,
I.\thinspace Minashvili$^{   7}$,
J.P.\thinspace Peigneux$^{  1}$,
V.\thinspace Romanovsky$^{   7}$,
N.\thinspace Russakovich$^{   7}$,
A.\thinspace Semenov$^{   7}$,
P.M.\thinspace Shagin$^{  5}$,
H.\thinspace Shimizu$^{ 10}$,
A.V.\thinspace Singovsky$^{ 1,5}$,
A.\thinspace Sobol$^{   5}$,
M.\thinspace Stassinaki$^{   2}$,
J.P.\thinspace Stroot$^{  6}$,
K.\thinspace Takamatsu$^{ 9}$,
T.\thinspace Tsuru$^{   8}$,
O.\thinspace Villalobos Baillie$^{   3}$,
M.F.\thinspace Votruba$^{   3}$,
Y.\thinspace Yasu$^{   8}$.
%% \end authorlist
}\end{center}

\begin{center}{\bf {{\bf Abstract}}}\end{center}

{
The reactions
$ pp \rightarrow p_{f} (X^0) p_{s}$, where $X^0$ is observed
decaying to  \etaetap and \etapetap,
have been studied at 450 GeV/c.
This is the first time that these channels have been observed in
central production and only the second time that
the \etapetap channel
has been observed
in any production
mechanism.
In the \etaetap channel there is evidence for the $f_0(1500)$ and
a peak at 1.95~GeV. The \etapetap channel shows
a peak at threshold which is compatible with having $J^{PC}~=~2^{++}$
and spin projection $J_Z$~=~0.
}
\bigskip
\bigskip
\bigskip
\bigskip\begin{center}{{Submitted to Physics Letters}}
\end{center}
%\newpage
\bigskip
\bigskip
\begin{tabbing}
aba \=   \kill
% $^\dag$ \> \small
% Deceased. \\
$^1$ \> \small
LAPP-IN2P3, Annecy, France. \\
$^2$ \> \small
Athens University, Physics Department, Athens, Greece. \\
%% $^3$ \> \small
%% Bergen University, Bergen, Norway. \\
$^3$ \> \small
School of Physics and Astronomy, University of Birmingham, Birmingham, U.K. \\
$^4$ \> \small
CERN - European Organization for Nuclear Research, Geneva, Switzerland. \\
$^5$ \> \small
IHEP, Protvino, Russia. \\
$^6$ \> \small
IISN, Belgium. \\
$^7$ \> \small
JINR, Dubna, Russia. \\
$^8$ \> \small
High Energy Accelerator Research Organization (KEK), Tsukuba, Ibaraki 305-0801,
Japan. \\
$^{9}$ \> \small
Faculty of Engineering, Miyazaki University, Miyazaki 889-2192, Japan. \\
$^{10}$ \> \small
RCNP, Osaka University, Ibaraki, Osaka 567-0047, Japan. \\
$^{11}$ \> \small
Oslo University, Oslo, Norway. \\
$^{12}$ \> \small
Faculty of Science, Tohoku University, Aoba-ku, Sendai 980-8577, Japan. \\
\end{tabbing}
\end{titlepage}
\setcounter{page}{2}
\bigskip
\par
The \etaeta, \etaetap and \etapetap channels are considered
to be promising places to look for glueballs since
it is thought likely that glueballs will decay
with the emission of
$\eta$s and $\eta^{\prime}$s~\cite{re:gers}.
Central production is proposed as a good place to search for
glueballs via Double Pomeron Exchange (DPE)~\cite{re:b}.
However, to date, only the \etaeta channel has been studied
in this production mechanism~\cite{na12etaeta}.
In fact the \etapetap channel has only ever been observed
once and that was by the VES experiment in the reaction
$\pi^-p \rightarrow  \eta^\prime \eta^\prime n$ where they
observed 14 events~\cite{VES}.
\par
In this paper a study is presented of the \etaetap and \etapetap
final states formed in the reaction
\begin{equation}
pp \rightarrow p_{f} (X^0) p_{s}
\label{eq:e}
\end{equation}
at 450~GeV/c.
The data come from the WA102 experiment
which has been performed using the CERN Omega Spectrometer,
the layout of which is
described in ref.~\cite{WADPT}.
\par
Reaction~(\ref{eq:e}), with $X^0$ being the \etaetap final state
has been isolated using the following decay modes:
\[
\begin{array}{cc}
\eta \rightarrow \gamma \gamma &\eta^\prime \rightarrow \eta \pi^+\pi^-, \eta
\rightarrow \gamma \gamma \\
\eta \rightarrow \gamma \gamma &\eta^\prime \rightarrow \eta \pi^+\pi^-, \eta
\rightarrow \pi^+\pi^-\pi^0 \\
\eta \rightarrow \pi^+\pi^-\pi^0 &\eta^\prime \rightarrow \eta \pi^+\pi^-, \eta
\rightarrow \gamma \gamma \\
\eta \rightarrow \gamma \gamma &\eta^\prime \rightarrow \rho^0(770) \gamma,
\rho^0(770) \rightarrow \pi^+\pi^- \\
\end{array}
\]
Other decay modes are possible but they have been found either
to have too much combinatorial background or have too few events
due to the small branching fraction of the $\eta^\prime$.
The above decay modes account for 26.5~\% of the total.
The $\rho^0(770)$ is observed decaying to $\pi^+\pi^-$
and is selected by requiring
0.7~$\leq$~M($\pi^+\pi^-$)~$\leq$~0.84~GeV.
\par
Fig.~\ref{fi:1}a) shows a scatter plot of $M(\gamma \gamma)$ against
$M(\eta \pi^+\pi^-)$ with $\eta \rightarrow \gamma \gamma$
which has been extracted from the sample of events having
two outgoing
central charged tracks and four $\gamma$s reconstructed in the GAMS-4000
calorimeter using momentum and energy balance.
A clear signal of the \etaetap channel can be observed.
Fig.~\ref{fi:1}e) shows the $\gamma \gamma$ mass spectrum if
the \etapipi is compatible with being an $\eta^\prime$
(0.9~$\leq$~M(\etapipi)~$\leq$~1.02~GeV) where a clear $\eta$ signal can
be observed.
Fig.~\ref{fi:1}f) shows the \etapipi mass spectrum if
the $\gamma \gamma$ is compatible with being an $\eta$
(0.45~$\leq$~M($\gamma \gamma$)~$\leq$~0.65~GeV) where a clear
$\eta^\prime$ signal can be observed.
The \etaetap final state has been selected using the mass cuts described above.
\par
Fig.~\ref{fi:1}b) shows a scatter plot of $M(\gamma \gamma)$ against
M($\eta \pi^+\pi^-)$ with $\eta \rightarrow \pi^+\pi^-\pi^0$
for the
sample of events having four
outgoing
central charged tracks and four $\gamma$s reconstructed in the GAMS-4000
calorimeter after imposing momentum and energy balance.
A clear signal of the \etaetap channel can be observed.
The \etaetap final state has been selected by requiring that the
0.45~$\leq$~M($\gamma \gamma$)~$\leq$~0.65~GeV and
0.9~$\leq$~M(\etapipi)~$\leq$~1.02~GeV.
Fig.~\ref{fi:1}c) shows a scatter plot of $M(\pi^+\pi^-\pi^0)$ against
$M(\eta \pi^+\pi^-)$ with $\eta \rightarrow \gamma \gamma$
for the
the sample of events having four
outgoing
central charged tracks and four $\gamma$s reconstructed in the GAMS-4000
calorimeter after imposing momentum and energy balance.
A signal of the \etaetap channel can be observed.
The \etaetap final state has been selected by requiring that the
0.5~$\leq$~M($\pi^+\pi^-\pi^0$)~$\leq$~0.6~GeV and
0.9~$\leq$~M(\etapipi)~$\leq$~1.02~GeV.
Fig.~\ref{fi:1}d) shows a scatter plot of $M(\gamma \gamma)$ against
$M(\rho^0(770) \gamma)$
for the
the sample of events having two
outgoing
central charged tracks and three $\gamma$s reconstructed in the GAMS-4000
calorimeter after imposing momentum and energy balance.
A signal of the \etaetap channel can be observed.
The \etaetap final state has been selected by requiring that the
0.45~$\leq$~M($\gamma \gamma$)~$\leq$~0.65~GeV and
0.9~$\leq$~M($\rho^0(770) \gamma$)~$\leq$~1.02~GeV.
\par
The background varies from 30 \% to 50 \% dependent on the
decay topology.
The total background in the combined \etaetap mass spectrum is
38 \% of the data.
In order to determine the effect of this background
events that
have the same final state particles but do not balance momentum have
been studied.
There is no signal for the $\eta$ or  $\eta^\prime$ in the
corresponding mass spectra but the distributions do represent the
background in the data quite well.
The effect of this background in the \etaetap mass spectrum
is a smooth distribution that reaches a maximum at 1.8 GeV.
\par
The resulting \etaetap mass spectra from each channel are
very similar and the combined mass spectrum is shown in
fig.~\ref{fi:2}a) and consists of 872 events.
The mass spectrum has a threshold enhancement and
a shoulder around 1.95 GeV.
In previous analyses of the \etaetap system by Crystal Barrel~\cite{cbetaetap}
and NA12~\cite{na12etaetap} this threshold has been interpreted as
being due to the $f_0(1500)$.
\par
In order to determine the spin of the \etaetap system,
an analysis has been
performed assuming that the \etaetap system is produced by the
collision of two particles (referred to as exchanged particles) emitted
by the scattered protons.
The z axis
is defined by the momentum vector of the
exchanged particle with the greatest four-momentum transferred
in the \etaetap centre of mass.
The y axis is defined
by the cross product of the two exchanged particles momenta
in the $pp$ centre of mass.
The two variables needed to specify the decay process were taken as the polar
and azimuthal angles ($\theta$, $\phi$) of the $\eta$ in the \etaetap
centre of mass relative to the coordinate system described above.
\par
Fig.~\ref{fi:2}c) shows the acceptance corrected
cos($\theta)$ distribution
for the $f_0(1500)$ region. The distribution is flat as would be expected
for a spin 0 particle.
%A similar study of the 1.9 GeV region also shows a flat distribution XXXXXXXX.
A fit has been performed to the
mass spectrum using
a  Flatt\'{e} like
formula~\cite{flatte} to describe the $f_0(1500)$, a
Breit-Wigner to describe the shoulder at 1.95~GeV
and a background
of the form
$a(m-m_{th})^{b}exp(-cm-dm^{2})$, where
$m$ is the
\etaetap
mass,
$m_{th}$ is the
\etaetap
threshold mass and
a, b, c, d are fit parameters.
The resulting fit is shown in fig.~\ref{fi:2}a) and gives
sheet II pole positions~\cite{sheet} for the $f_0(1500)$ of
\begin{tabbing}
00000aaaa\=adfsfsf99ba \=Mas  == 1224 pm0 \=i12\=2400 \=pi \=1200000  \=MeV
\kill
\>$f_0(1500)$ \>M $ \; = \;$(1515$\; \pm\; $12)\>$-i$\>($ \;
\;55$\>$\pm$\>12)\>MeV\\
\end{tabbing}
These parameters are consistent with the PDG~\cite{PDG98} values for the
$f_0(1500)$.
For the peak at 1.95~GeV the fit gives
\begin{tabbing}
aaaa\=adfsfsf99ba \=Mas \= == \=1224 \=pm \=1200 \=MeVswfw, \=gaa \=  == \=1224
\=pm \=1200  \=MeV   \kill
\>         \>M \>=\>1934\>$\pm$\>16\>MeV,\>$\Gamma$\>=\>141\>$\pm$\>41\>MeV
\end{tabbing}
These parameters are similar to those found in the
\etaetap final state by the NA12 experiment~\cite{na121910} which they
claimed may have an exotic nature.
The PDG~\cite{PDG98} have listed this observation with a state seen
with similar mass and width in the $\omega \omega$ final state which
has been found to have $J^{PC}$~=~$2^{++}$.
This state is called the $X(1910)$ by the PDG~\cite{PDG98}.
The background distribution found from the fit is
similar in size and shape to that which was determined by using events that
do not balance momentum described above.
\par
In order to try to learn more about the spin of the peak at 1.95 GeV,
the \etaetap mass spectrum has been fitted in four intervals of
cos($\theta$) and the number of events in each bin determined from the fit.
The resulting cos($\theta$) distribution is shown in fig.~\ref{fi:2}d).
Superimposed on the distribution are the three spin hypotheses which
best describe the data. The solid curve represents a $J^{PC}$~=~$0^{++}$
state, the dashed curve a
$J^{PC}$~=~$1^{-+}$ with spin projection $|J_Z|$~=1 and the
dotted curve a
$J^{PC}$~=~$2^{++}$ with spin projection $|J_Z|$~=2. All other
spin projections can be discounted. It should be noted that no evidence
has previously been found in central production for a resonance
with spin projection $|J_Z|$~=~2.
Further information will come from an analysis of the
centrally produced $\omega \omega$ system which is in progress.
\par
It is possible to decrease the $\chi^2/NDF$ of the fit to the mass
spectrum from 16/15 to 10/12 by introducing a third
Breit-Wigner in the 2.4~GeV region. The resulting parameters for this
Breit-Wigner are $M$~=~2369~$\pm$~10~MeV, $\Gamma$~=~52~$\pm$~31~MeV.
If this were interpreted as a resonance it could be due to the
$f_2(2340)$ which has been observed in the centrally produced
$\phi \phi$ final state~\cite{pi4papr}.
\par
In previous analyses a study has been made of how different
resonances are produced as a function of the
parameter $dP_T$, which is the difference
in the transverse momentum vectors of the two exchange
particles~\cite{WADPT}, and of
the azimuthal angle $\phi$ which is defined as the angle between the $p_T$
vectors of the two outgoing protons.
\par
In order to learn more about the resonances produced in the \etaetap channel
a study has been made of the $dP_T$ and $\phi$ dependences. These
dependences are consistent with the threshold region
being due to the $f_0(1500)$~\cite{pipikkpap}.
\par
Correcting for the unseen decay modes and the effects of the detector
the branching ratio of the $f_0(1500)$ to \etaetap/$\pi \pi$ is
determined to be 0.095~$\pm$~0.026, consistent with the value
that is derived from the PDG~\cite{PDG98} of 0.066~$\pm$~0.033.
\par
After taking into account the background, correcting for
geometrical acceptances, detector efficiencies,
losses due to cuts
and unseen decay modes,
the cross-section for
the \etaetap channel at $\sqrt s$~=~29.1~GeV in the
$x_F$ interval
$|x_F| \leq 0.2$ is $\sigma$(\etaetap)~=~145~$\pm$~18~nb.
Above the \etapetap threshold the cross section is 53~$\pm$~10~nb.
\par
The remainder of this paper describes the \etapetap channel.
Reaction~(\ref{eq:e}), with $X^0$ being the \etapetap final state
has been isolated using the following decay modes:
\[
\begin{array}{cc}
\eta^\prime \rightarrow \rho^0(770) \gamma &\eta^\prime \rightarrow \rho^0(770)
\gamma \\
\eta^\prime \rightarrow \rho^0(770) \gamma &\eta^\prime \rightarrow \eta
\pi^+\pi^-, \eta \rightarrow \gamma \gamma \\
\eta^\prime \rightarrow \eta \pi^+\pi^-, \eta \rightarrow \gamma \gamma
&\eta^\prime \rightarrow \eta \pi^+\pi^-, \eta \rightarrow \gamma \gamma \\
\end{array}
\]
Other decay modes are possible but they have been found either
to have too much combinatorial background or have too few events
due to the small branching fraction of the $\eta^\prime$.
The above decay modes account for 22.4~\% of the total.
\par
Fig.~\ref{fi:3}a) shows a scatter plot of $M(\rho^0(770) \gamma)$ against
$M(\rho^0(770) \gamma)$
which has been extracted from the sample of events having
four outgoing
central charged tracks and two $\gamma$s reconstructed in the GAMS-4000
calorimeter using momentum and energy balance.
A clear signal of the \etapetap channel can be observed.
Fig.~\ref{fi:3}b) shows the $\rho^0(770) \gamma$ mass spectrum if
the other $\rho^0(770) \gamma$ mass is compatible with being an $\eta^\prime$
(0.9~$\leq$~$M(\rho^0(770)\gamma)$~$\leq$~1.02~GeV)
where a clear $\eta^\prime$ signal can
be observed.
The \etapetap final state has been selected using the mass cuts described
above.
\par
Fig.~\ref{fi:3}c) shows a scatter plot of $M(\rho^0(770) \gamma)$ against
$M(\eta \pi^+\pi^-)$ with the $\eta$ decaying to $\gamma \gamma$
which has been extracted from the sample of events having
four outgoing
central charged tracks and three $\gamma$s reconstructed in the GAMS-4000
calorimeter using momentum and energy balance.
A clear signal of the \etapetap channel can be observed.
Fig.~\ref{fi:3}d) shows the $\rho^0(770) \gamma$ mass spectrum if
the $\eta \pi^+ \pi^-$ mass is compatible with being an $\eta^\prime$
(0.9~$\leq$~M(\etapipi)~$\leq$~1.02~GeV)
and fig~\ref{fi:3}e) shows the \etapipi mass spectrum if the
$\rho^0(770) \gamma$ mass is compatible with being an $\eta^\prime$.
In both cases a
clear $\eta^\prime$ signal can
be observed.
The \etapetap final state has been selected using the mass cuts described
above.
\par
Fig.~\ref{fi:3}f) shows a scatter plot of $M(\eta \pi^+\pi^-)$ against
$M(\eta \pi^+ \pi^-)$ with both $\eta$s decaying to $\gamma \gamma$
which has been extracted from the sample of events having
four outgoing
central charged tracks and four $\gamma$s reconstructed in the GAMS-4000
calorimeter using momentum and energy balance.
A clear signal of the \etapetap channel can be observed.
Fig.~\ref{fi:3}g) shows the \etapipi mass spectrum if
the other \etapipi mass is compatible with being an $\eta^\prime$
(0.9~$\leq$~M(\etapipi)~$\leq$~1.02~GeV) where a clear $\eta^\prime$ signal can
be observed.
The \etapetap final state has been selected using the mass cuts described
above.
\par
The background varies from 23 \% to 55 \% dependent on the
decay topology.
The total background in the combined \etaetap mass spectrum is
36 \% of the data.
In order to determine the effect of this background
events that
have the same final state particles but do not balance momentum have
been studied.
There is no signal for the $\eta^\prime$ in the
corresponding mass spectra but the distributions do represent the
background in the data quite well.
The effect of this background in the \etapetap mass spectrum
is a smooth distribution that reaches a maximum at 2.2 GeV.
\par
The resulting \etapetap mass spectra from each channel are
very similar and the combined mass spectrum consists of 166 events and
is shown in
fig.~\ref{fi:4}a) in 40 MeV bins and in fig.~\ref{fi:4}b) in 80 MeV bins.
The mass spectrum has a peak around 2~GeV.
\par
Fig.~\ref{fi:4}c) shows the acceptance corrected
cos($\theta)$ distribution
for the 2~GeV region. The distribution is not flat.
Superimposed on the distribution is the result
of a fit to the distribution of the form
$\alpha+ \beta(3/2cos ^2 \theta -1/2)^2$
representing what would be expected
for a spin 2 particle with spin projection $J_Z$~=~0 and a flat
background. The fit well reproduces the data and gives
$\alpha$~=~1.5 and $\beta$~=~9.0.
\par
A fit has been performed to the
mass spectrum using
a
Breit-Wigner to describe the 2~GeV region
and a background
of the form
$a(m-m_{th})^{b}exp(-cm-dm^{2})$, where
$m$ is the
\etapetap
mass,
$m_{th}$ is the
\etapetap
threshold mass and
a, b, c, d are fit parameters.
The fit results in $M$~=~2007~$\pm$~24 MeV, $\Gamma$~=~90~$\pm$~43~MeV.
Based on the mass and width of this state and taking into account
the \etapetap threshold this peak could be described
as being due to the $X(1910)$ observed in the
\etaetap channel. However the spin projection
found for the $X(1910)$ is completely different.
\par
The broad distribution, described as the background above,
is compatible in shape with what would be expected from the
$f_2(1950)$ observed decaying to $f_2(1270) \pi \pi$
which also has possible $\phi \phi$ and $K^* \overline K^*$
decay modes~\cite{pi4papr}.
In order to see if the $f_2(1950)$ does have an \etapetap
decay mode
a study has been made of the $dP_T$ and $\phi$ dependences of the
\etapetap  channel.
The fraction of \etapetap events
has been calculated for
$dP_T$$\leq$0.2 GeV, 0.2$\leq$$dP_T$$\leq$0.5 GeV and $dP_T$$\geq$0.5 GeV and
gives
0.10~$\pm$~0.03, 0.44~$\pm$~0.05 and 0.46~$\pm$~0.05 respectively.
This results in a ratio of production at small $dP_T$ to large $dP_T$ of
0.22~$\pm$~0.07.
The azimuthal angle ($\phi$) between the $p_T$
vectors of the two protons
is shown in
fig.~\ref{fi:4}d).
Both the $dP_T$ and $\phi$ distributions are different to what
was observed for the $f_2(1950)$~\cite{pi4papr} and hence would suggest
that the \etapetap channel is not mainly due to the $f_2(1950)$.
\par
After taking into account the background, correcting for
geometrical acceptances, detector efficiencies,
losses due to cuts,
and unseen decay modes,
the cross-section for
the \etapetap channel at $\sqrt s$~=~29.1~GeV in the
$x_F$ interval
$|x_F| \leq 0.2$ is $\sigma$(\etapetap)~=~146~$\pm$~24~nb.
\par
In summary,
a study of the \etaetap and \etapetap channels has been performed
for the first time in central production. The \etapetap channel
has been observed with more than a factor of 10 times the statistics
of the only other previous observation
of this channel in any production
mechanism~\cite{VES}.
In the \etaetap channel there is evidence for the $f_0(1500)$ and
a peak at 1.95~GeV. The \etapetap channel shows
a peak at threshold which is compatible with having
$J^{PC}~=~2^{++}$
and spin projection $J_Z$~=~0.
\begin{center}
{\bf Acknowledgements}
\end{center}
\par
This work is supported, in part, by grants from
the British Particle Physics and Astronomy Research Council,
the British Royal Society,
the Ministry of Education, Science, Sports and Culture of Japan
(grants no. 07044098 and 1004100), the French Programme International
de Cooperation Scientifique (grant no. 576)
and
the Russian Foundation for Basic Research
(grants 96-15-96633 and 98-02-22032).

\clearpage
{ \large \bf Figures \rm}
\begin{figure}[h]
\caption{
Selection of the \etaetap final state.
a) $M(\gamma \gamma)$ versus $M(\eta \pi^+\pi^-)$ with
$\eta \rightarrow \gamma \gamma$,
b) $M(\gamma \gamma)$ versus
M($\eta \pi^+\pi^-)$ with $\eta \rightarrow \pi^+\pi^-\pi^0$,
c) $M(\pi^+\pi^-\pi^0)$ versus
$M(\eta \pi^+\pi^-)$ with $\eta \rightarrow \gamma \gamma$,
d) $M(\gamma \gamma)$ versus
$M(\eta \pi^+\pi^-)$ with $\eta \rightarrow \rho^0(770) \gamma$,
e) M($\gamma \gamma$) if
(0.9~$\leq$~M(\etapipi)~$\leq$~1.02~GeV) and
f) shows the M(\etapipi) if
(0.45~$\leq$~M($\gamma \gamma$)~$\leq$~0.65~GeV).
}
\label{fi:1}
\end{figure}
\begin{figure}[h]
\caption{
a) and b) the \etaetap mass spectrum with fit described in the text.
The cos($\theta)$ distribution a) for the $f_0(1500)$ region and
b) for the 1.95~GeV region.
}
\label{fi:2}
\end{figure}
\begin{figure}[h]
\caption{
Selection of the \etapetap final state.
a) M($\rho^0(770)\gamma )$ versus M($\rho^0(770) \gamma)$,
b) the M($\rho^0(770) \gamma)$ distribution if the other
M($\rho^0(770) \gamma)$ distribution is compatible with being an $\eta^\prime$.
c) M(\etapipi ) versus M($\rho^0(770) \gamma)$,
d) the M($\rho^0(770) \gamma)$ distribution if
M(\etapipi) distribution is compatible with being an $\eta^\prime$.
e) the M(\etapipi) distribution if
M($\rho^0(770)\gamma)$ distribution is compatible with being an $\eta^\prime$.
f) M(\etapipi ) versus M(\etapipi) and
g) the M(\etapipi) distribution if the other M(\etapipi) is
compatible with being an $\eta^\prime$.
}
\label{fi:3}
\end{figure}
\begin{figure}[h]
\caption{a) and b) the \etapetap mass distribution with fit
described in the text.
c) The cos($\theta)$ distribution for the 1.9 $-$ 2.1 GeV region.
d) the $\phi$ distribution for the \etapetap channel.
}
\label{fi:4}
\end{figure}
\begin{center}
\epsfig{figure=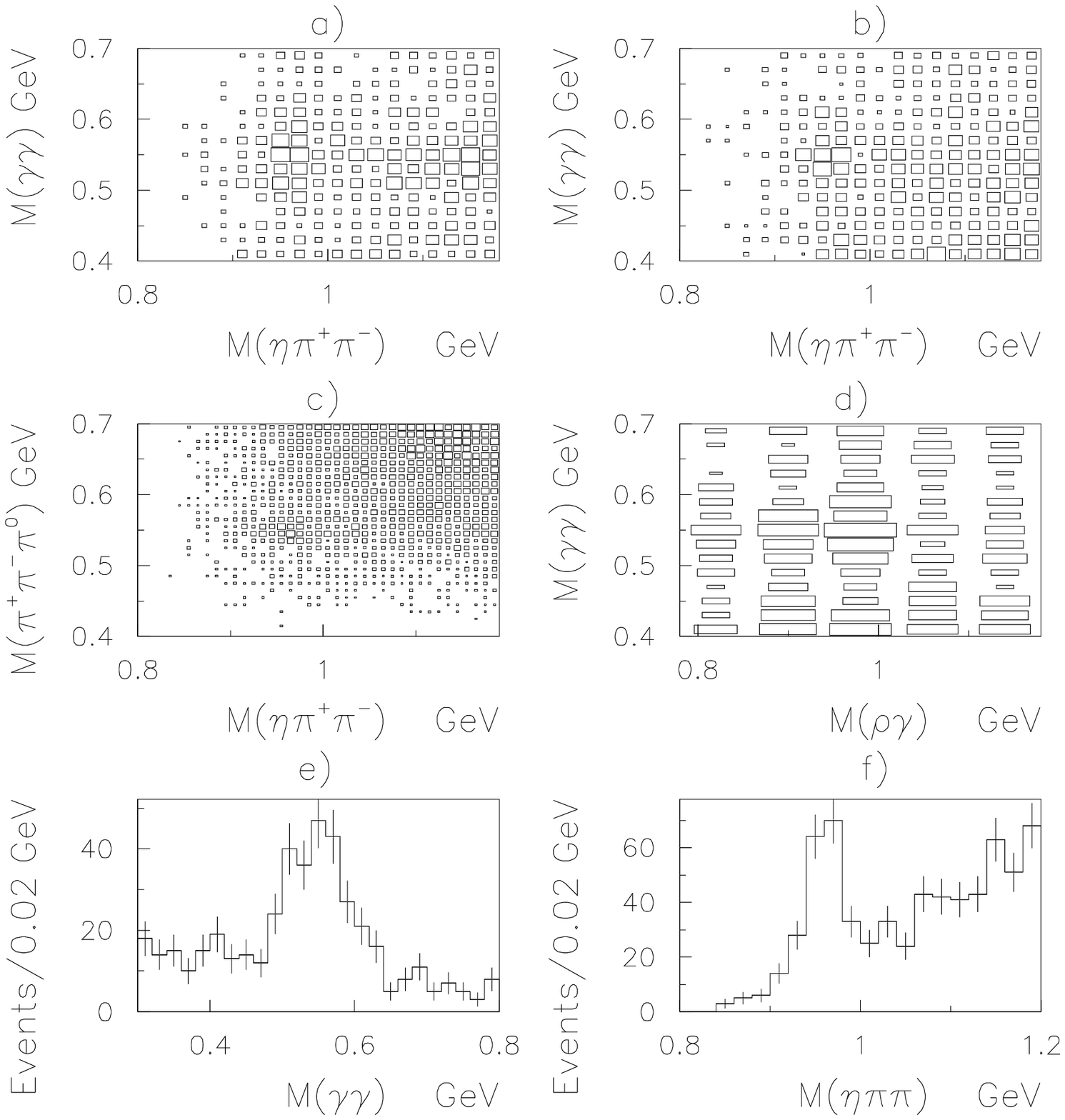,height=22cm,width=17cm}
\end{center}
\begin{center} {Figure 1} \end{center}
\newpage
\begin{center}
\epsfig{figure=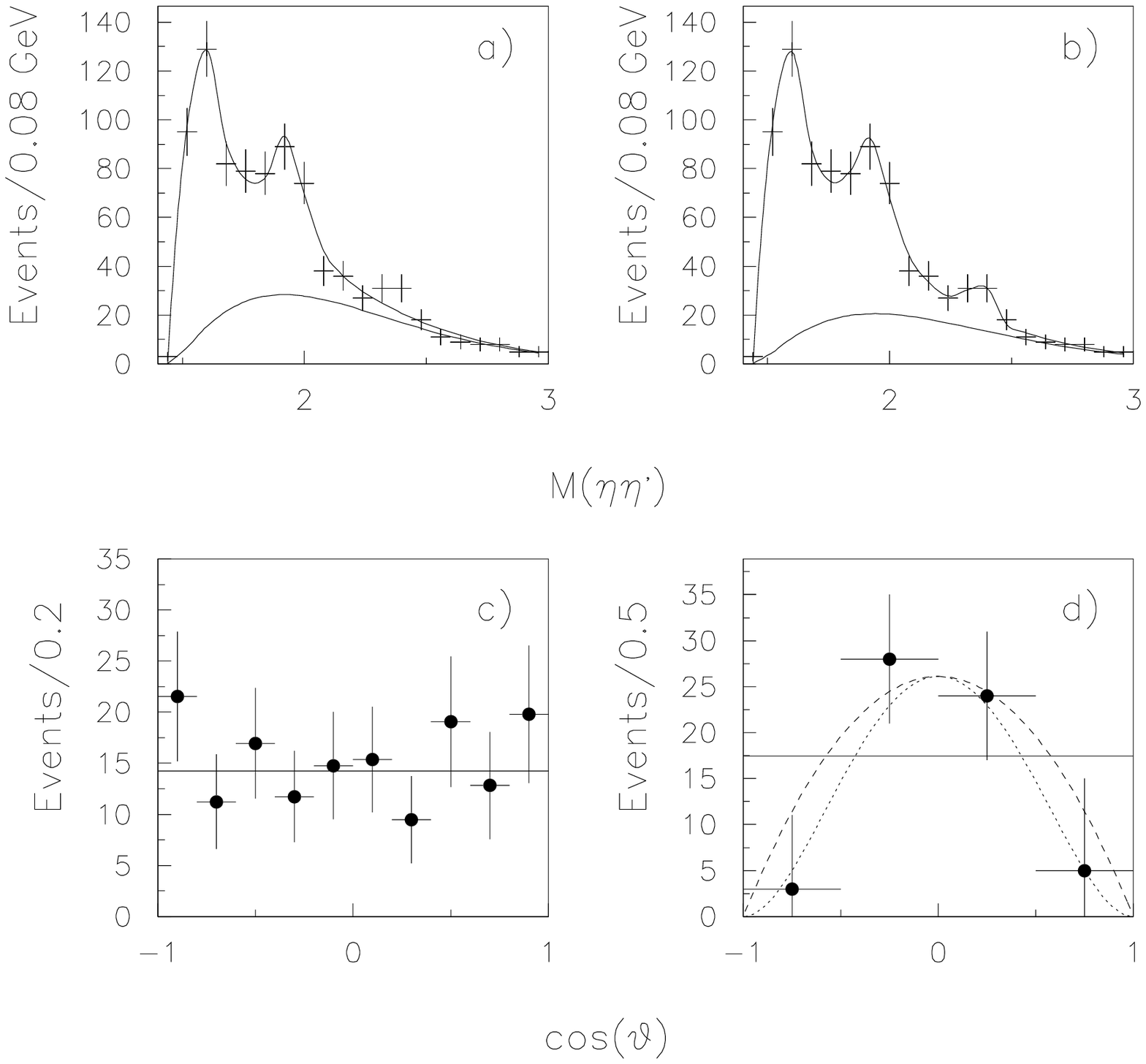,height=22cm,width=17cm}
\end{center}
\begin{center} {Figure 2} \end{center}
\newpage
\begin{center}
\epsfig{figure=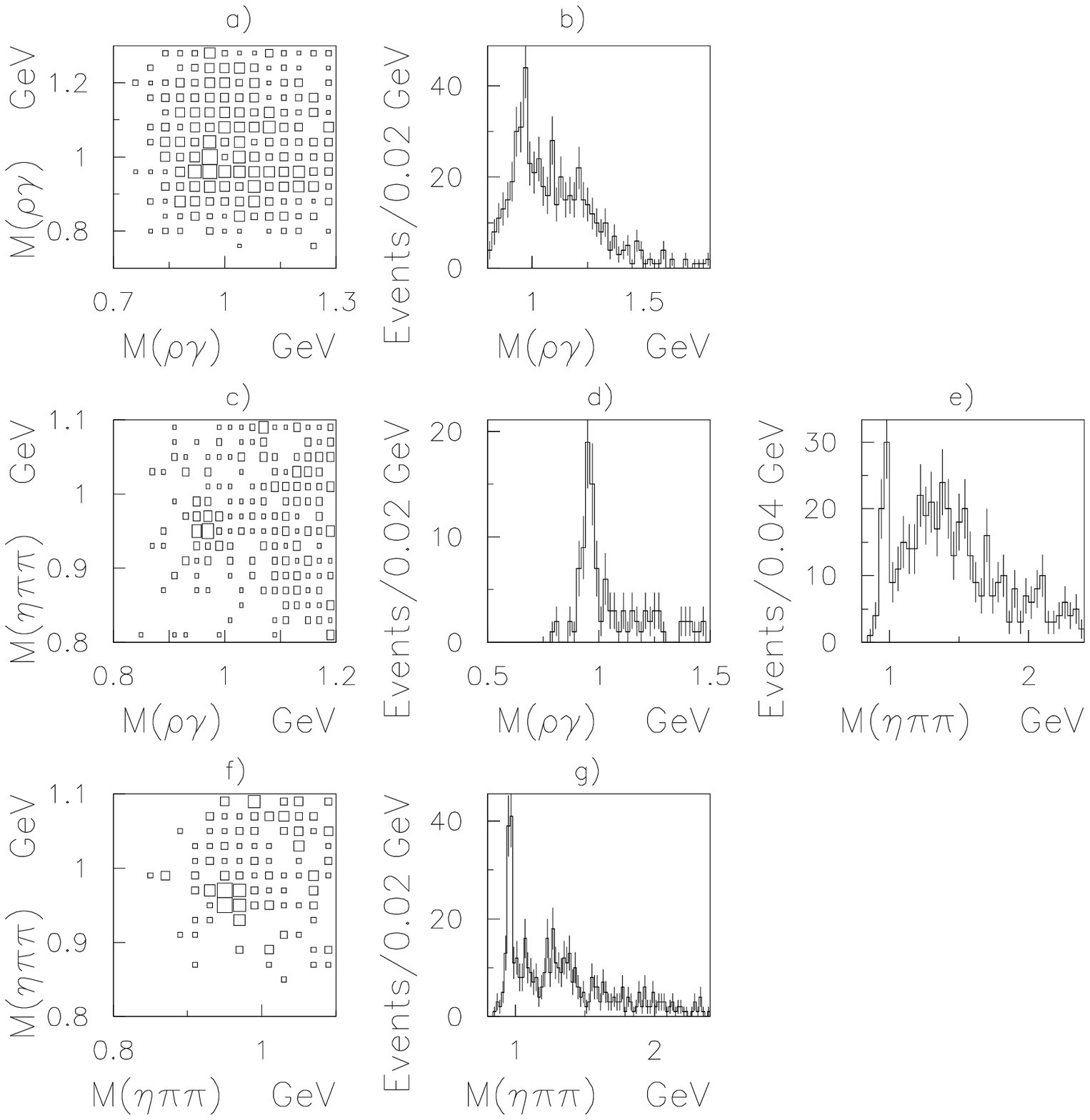,height=22cm,width=17cm}
\end{center}
\begin{center} {Figure 3} \end{center}
\newpage
\begin{center}
\epsfig{figure=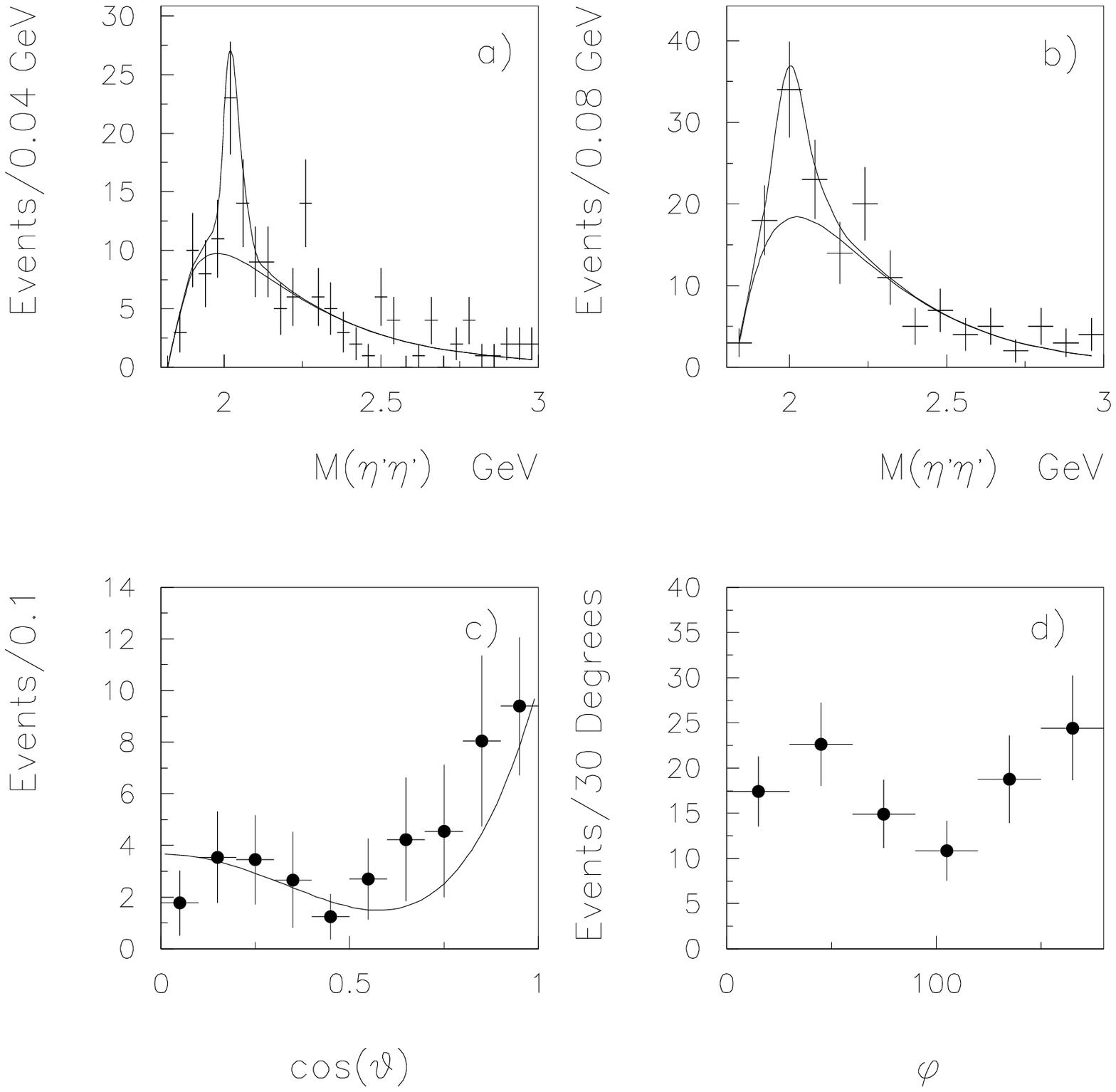,height=22cm,width=17cm}
\end{center}
\begin{center} {Figure 4} \end{center}

\bigskip
\newpage
\begin{thebibliography}{99}
\bibitem{re:gers}
S. S. Gershtein {\em et al.,} Zeit. Phys. {\bf C24} (1984) 305; \\
R. Akhoury and J.M. Frere, Phys. Lett. {\bf B220} (1989) 258.
\bibitem{re:b}
D. Robson {\em et al.,} Nucl. Phys. {\bf B130 } \rm (1977) 328; \\
F. E. Close, Rep. Prog. Phys. {\bf 51} (1988) 833.
\bibitem{na12etaeta}
D. Alde {\em et al.,} Phys. Lett. {\bf B201} (1988) 160; \\
A. Singovski, Il Nouvo Cimento {\bf A107} (1993) 1911.
\bibitem{VES}
G. M. Beladidze {\em et al.,} Zeit. Phys. {\bf C57 } (1992) 13.
\bibitem{WADPT}
D. Barberis {\em et al.,} Phys. Lett. {\bf B397 } \rm (1997) 339.
\bibitem{cbetaetap}
C. Amlser {\em et al.,} Phys. Lett. {\bf B353} (1995) 571.
\bibitem{na12etaetap}
F. Binon {\em et al.,} Il Nouvo Cimento {\bf A80} (1994) 363.
\bibitem{flatte}
S.M. Flatt\'{e}, Phys. Lett. {\bf B38} (1972) 232.
\bibitem{sheet}
D. Morgan, Phys. Lett. {\bf B51} (1974) 71.
\bibitem{PDG98}
Particle Data Group, European Physical Journal {\bf C3} (1998) 1.
\bibitem{na121910}
D. Alde {\em et al.,} Phys. Lett. {\bf B216 } (1989) 447.
\bibitem{pi4papr}
D. Barberis {\em et al.,} In preparation.
\bibitem{pipikkpap}
D. Barberis {\em et al.,} Phys. Lett. {\bf B462} (1999) 462.
\end{thebibliography}
\end{document}